\documentclass[12pt,english,floatfix,nofootinbib,superscriptaddress,aps,prd,preprint]{revtex4}
\usepackage[utf8]{inputenc}
\usepackage{float}
\usepackage{array}
\usepackage{lipsum}
\usepackage{graphicx}
\usepackage{amsmath,amsthm,amsfonts,amssymb}
\usepackage{graphicx}
\usepackage[english]{babel}
\usepackage{color}
\usepackage{tensor}
\usepackage{esint}
\usepackage[dvips]{epsfig}
\usepackage[dvips]{graphicx}
\usepackage{float}
\usepackage{units}
\usepackage{textcomp}
\usepackage{mathrsfs}
\usepackage{amsmath}
\usepackage[makeroom]{cancel}
\usepackage{amssymb}
\usepackage{amsbsy}
\usepackage{amsfonts}
\usepackage{amssymb,mathrsfs,xcolor}
\usepackage{esint}
\usepackage{array}
\usepackage{graphicx}

\usepackage{hyperref}
\hypersetup{
    colorlinks,
    citecolor=blue,
    filecolor=green,
    linkcolor=purple,
    urlcolor=red,
}

\usepackage{slashed}

\newcommand{\ie}{\begin{equation}}

\newcommand{\se}{\begin{eqnarray}}
\newcommand{\ff}{\end{eqnarray}}

\usepackage{hyperref}
\hypersetup{colorlinks,breaklinks,
			citecolor=[rgb]{0,0.0,1.0},
            urlcolor=[rgb]{0.0,0.0,1.0},
            linkcolor=[rgb]{0,0.5,0.9}}

\begin{document}

\title{Generalized Chern-Pontryagin models}

\author{J. R. Nascimento}
\email{jroberto@fisica.ufpb.br}
\affiliation{Departamento de Física, Universidade Federal da Paraíba, Caixa Postal 5008, 58051-970, João Pessoa, Paraíba,  Brazil} 

\author{A. Yu. Petrov}
\email{petrov@fisica.ufpb.br}
\affiliation{Departamento de Física, Universidade Federal da Paraíba, Caixa Postal 5008, 58051-970, João Pessoa, Paraíba,  Brazil} 

\author{P. J. Porfírio}
\email{pporfirio@fisica.ufpb.br}
\affiliation{Departamento de Física, Universidade Federal da Paraíba, Caixa Postal 5008, 58051-970, João Pessoa, Paraíba,  Brazil} 

\author{Ramires N. da Silva}
\email{rns2@academico.ufpb.br }
\affiliation{Departamento de Física, Universidade Federal da Paraíba, Caixa Postal 5008, 58051-970, João Pessoa, Paraíba,  Brazil}




\date{\today}

\begin{abstract}

We formulate a new class of modified gravity models, that is, generalized four-dimensional Chern-Pontryagin models, whose action is characterized by an arbitrary function of the Ricci scalar $R$ and the Chern-Pontryagin topological term ${}^*RR$, i.e., $f(R, {}^*RR)$. Within this framework, we derive the gravitational field equations and solve them for the particular models, $f(R, {}^*RR)=R+\beta ({}^*RR)^2$ and $f(R, {}^*RR)=R+\alpha R^2+\beta ({}^*RR)^2$, considering two ansatzes: the slowly rotating metric and first-order perturbations of G\"{o}del-type metrics. For the former, we find a first-order correction to the frame-dragging effect boosted by the parameter $L$, which characterizes the departures from general relativity results. For the latter, G\"{o}del-type metrics hold unperturbed, for specific sort of perturbed metric functions. We conclude this paper by displaying that generalized four-dimensional Chern-Pontryagin models admit a scalar-tensor representation, whose explicit form presents two scalar fields: $\Phi$, a dynamical degree of freedom, while the second, $\vartheta$, a non-dynamical degree of freedom. In particular, the scalar field $\vartheta$ emerges coupled with the Chern-Pontryagin topological term ${}^*RR$, i.e., $\vartheta {}^*RR$, which is nothing more than Chern-Simons term.


\end{abstract}

\maketitle


\section{Introduction}

Studies of gravitational phenomena are one of the most important research lines in contemporary physics. Such an interest is inspired in part by important recent discoveries, such as the late-time accelerating expansion of the Universe \cite{Riess}, detection of gravitational waves \cite{LIGO} and obtaining an image of the shadow of a black hole \cite{Akiyama}.  Within this context, modified gravity models have played a pivotal role since they can tackle open questions in gravitational physics or, in the cosmological scenario, provide alternative answers to $\Lambda$CDM without making use of the addition of a certain amount of unseen exotic matter/energy \cite{Clifton,Ourbook,Olmo2011}.

As it is known, various extensions of Einstein's gravity have been proposed -- those including modification in the pure gravity sector \cite{Sot,DeFelice,Noj,Nojiri:2017ncd} and others involving extra fields, either scalar \cite{Brans,Horndeski, Koba}, vector \cite{KosGra,Jacob,Del} or even tensor ones (that is, so-called bimetric gravity). One of such extensions, that has attracted a strong interest in recent years, is the four-dimensional Chern-Simons modified gravity (CSMG) originally proposed in \cite{JaPi}. In essence, the CSMG is the theory whose Lagrangian includes, besides of the scalar curvature, the Chern-Pontryagin topological term ${}^*RR\equiv\frac{1}{2}\epsilon^{\mu\nu\lambda\rho}R_{\mu\nu\alpha\beta}R_{\lambda\rho}^{\phantom{\lambda\rho}\alpha\beta}$, multiplied by an extra scalar $\vartheta$ called the Chern-Simons (CS) coefficient, i.e., the $\vartheta{}^*RR$ term (also called the CS term) is added. This term, being the natural four-dimensional extension of the 3-dimensional gravitational Chern-Simons term \cite{DJT}, displays numerous interesting properties. First, it can break parity explicitly, depending on the form of $\vartheta$; if $\vartheta$ is even under parity transformation, once ${}^*RR$ is odd under parity transformation. Also, it can break Lorentz symmetry at the action level, since one can rewrite the CS term as an interaction term between a background vector $v_{\mu}\equiv \partial_{\mu}\vartheta$ and the topological Chern-Simons current $K^{\mu}$. Second, it was checked that CSMG shares some solutions with the general relativity (GR), for example, the Schwarzschild metric. Conversely, as a counterexample, one can mention the Kerr metric which is not a solution of CSMG, as discussed in \cite{Yunes,Konno,Alexander:2009tp}. In the further generalization of the CSMG, the so-called dynamical CSMG (DCSMG), the kinetic term for $\vartheta$ is introduced.

A detailed discussion of the consistency of various known metrics within the DCSMG is presented in \cite{Grumiller}, and various issues related to gravitational waves in metric DCSMG were studied in \cite{Nojiri:2019nar,Nojiri:2020pqr}. A metric-affine extension of the DCSMG with projective invariance has been developed \cite{Boudet}, and propagation of gravitational waves within this theory has also been investigated \cite{Boudet2, Bombacigno}. The Chern-Pontryagin topological invariant also appears naturally in the context of string/M-theory, specifically, compactifying $D=11$ M-theory down to $D=4$ effective gravity theories. In particular, one can cite the Starobinsky-Bel-Robinson (SBR) model \cite{Ketov:2022lhx,Ketov:2022zhp}, whose  action involves quadratic Chern-Pontryagin terms, $({}^*RR)^2$, its cosmological aspects were studied in \cite{Ketov:2022lhx,Ketov:2022zhp}. Another interesting string-inspired model closely related to SBR model is the Einstein-Grisaru-Zanon effective gravity theory which has been investigated in \cite{CamposDelgado:2024jst}.

One more motivation for studying the modified gravity theories, especially those ones involving the ${}^*RR$ term and its generalizations, is related with the possibility of breaking of Lorentz and/or CPT symmetries suggested already in the seminal papers \cite{KosSam}, where this possibility has been justified by string theory (further, other motivations for the possibility of Lorentz-CPT symmetry breaking, based on loop quantum gravity, minimal length hypothesis, space-time noncommutativity, etc. were introduced), therefore, various Lorentz-CPT breaking extensions of gravity can be defined (for a review of such extensions, see \cite{KosLi}). The simplest known example of the Lorentz-CPT breaking term in gravity is just the four-dimensional gravitational CS term \cite{JaPi}. Already in \cite{KosGra} possibilities to generalize this term, and hence to obtain higher-order Lorentz-CPT breaking terms in gravity were mentioned. Therefore, the natural idea consists in studying of gravity models involving additive terms of the form $({}^*RR)^n$ which, first, generalize the gravitational CS term, second, break the CPT symmetry for odd values of $n$, representing the most natural Riemannian CPT-breaking gravity models (we note that there are also non-Riemannian CPT-breaking theories \cite{Li, Porf1, Li2, Porf2, Li3, Li4, Porf3, Porf4, Rao, Porf5, Porf6}). 

In this paper, driven by $f(R)$ theories, we propose a more generic class of theories called generalized Chern-Pontryagin models, whose action, instead of being linear or quadratic in ${}^*RR$, displays a generic functional dependency on this object and also the Ricci scalar. Certainly, this will allow for a richer framework to explore, among other issues, new parity-violating solutions of the modified field equations in the model. At the same time, many well-known metrics, including Schwarzschild and G\"{o}del ones, actually, all metrics fulfilling the Pontryagin constraint ${}^*RR=0$, can be easily shown to be simultaneous solutions of GR and generalized Chern-Pontryagin theories. Then, it is interesting to examine the influence of the generalized Chern-Pontryagin term under small perturbations of these metrics within the modified field equations. Nonetheless, our main goal is to provide a natural mechanism to generate the four-dimensional CS term, which was originally proposed in analogy to electrodynamics \cite{JaPi}.

The structure of the paper looks like follows. In  Section II, we write down the classical action of the theory and obtain the corresponding equations of motion. In Section III, we study the consistency of slowly rotating black hole solutions and first-order perturbations of ST-homogeneous G\"{o}del-type metrics in the modified theory. In Section IV, we provide a mechanism to generate the Chern-Simons term. Finally, in Section V we discuss our results.

\section{Generalized Chern-Pontryagin 
gravity models}

We start this section by defining the action of the generalized Chern-Pontryagin gravity models, namely, 
\begin{equation}
    S=\frac{1}{2\kappa^2}\int d^{4}x\sqrt{-g}\, f(R,\,^{*}RR)+\int d^{4}x\sqrt{-g}\,\mathcal{L}_{m}(g_{\mu\nu},\Psi),
    \label{eq1}
\end{equation}
where $f$ is an arbitrary function of the Ricci scalar and the topological Chern-Pontryagin term $\,^{*}RR\equiv \frac{1}{2}\epsilon^{\mu\nu\alpha\beta}R^{\lambda}_{\;\; \tau\mu\nu}R^{\tau}_{\;\; \lambda\alpha\beta}$, $\mathcal{L}_{m}$ stands for the matter Lagrangian and $\Psi$ symbolically represents the matter sources. It is clear that this model in certain cases, namely, when the function $f(R,\,^{*}RR)$ includes odd degrees of the ${}^*RR$, breaks parity similarly to the CSMG (cf. \cite{JaPi}).

By varying the action (\ref{eq1}) with respect to the metric, we find the field equations, 
\begin{eqnarray}
    \nonumber\kappa^2 T_{\mu\nu}^{(m)}&=&f_{R}R_{\mu\nu}-\frac{f}{2}g_{\mu\nu}+g_{\mu\nu}\Box f_{R}-\nabla_{(\mu}\nabla_{\nu)}f_{R}+4\bigg(\,^{*}R^{\tau\,\,\,\,\,\,\,\,\,\,\lambda}_{\,\,\,(\mu\nu)}\nabla_{\lambda}\nabla_{\tau}f_{\,^{*}RR}+\\
&+&\epsilon^{\lambda\beta\gamma}_{\,\,\,\,\,\,\,\,\,\,(\nu}\nabla_{|\gamma|}R_{\mu)\beta}\nabla_{\lambda}f_{\,^{*}RR}\bigg)+\frac{1}{2}g_{\mu\nu}\,^{*}RR f_{\,^{*}RR},
    \label{fe}
\end{eqnarray}
where we have defined the following quantities: $f_{R}=\frac{\partial f}{\partial R}$, $f_{\,^{*}RR}=\frac{\partial f}{\partial (\,^{*}RR)}$ and $\Box\equiv g^{\mu\nu}\nabla_{\mu}\nabla_{\nu}$. Furthermore, 
\begin{equation}
T_{\mu\nu}^{(m)}=-\frac{2}{\sqrt{-g}}\frac{\delta(\sqrt{-g}\mathcal{L}_{m})}{\delta g^{\mu\nu}}
\end{equation}
represents the stress-energy tensor of the matter sources.

For our purposes, it should be noted that the field equations (\ref{fe}) can be rewritten in a more convenient way, namely,
\begin{equation}
G_{\mu\nu}=\kappa^2_{eff}T_{\mu\nu}^{(m)}+T_{\mu\nu}^{eff},
    \label{eq44}
\end{equation}
where $\kappa^2_{eff}=\dfrac{\kappa^2}{f_{R}}$, and
\begin{equation}
    T_{\mu\nu}^{eff}=\frac{1}{f_{R}}\left(-\frac{1}{2}f_{R}Rg_{\mu\nu}+\frac{f}{2}g_{\mu\nu}-g_{\mu\nu}\Box f_{R}+\nabla_{(\mu}\nabla_{\nu)}f_{R}-C_{\mu\nu}\right)
\end{equation}
is the effective stress-energy tensor. The $C$-tensor above is defined below
\begin{equation}
C_{\mu\nu}=4\bigg(\,^{*}R^{\tau\,\,\,\,\,\,\,\,\,\,\lambda}_{\,\,\,(\mu\nu)}\nabla_{\lambda}\nabla_{\tau}f_{\,^{*}RR}+\epsilon^{\lambda\beta\gamma}_{\,\,\,\,\,\,\,\,\,\,(\nu}\nabla_{|\gamma|}R_{\mu)\beta}\nabla_{\lambda}f_{\,^{*}RR}\bigg)+\frac{1}{2}g_{\mu\nu}\,^{*}RR f_{\,^{*}RR}.
\label{ctensor}
\end{equation}
 Note however that the field equations involve higher-order derivative terms whose presence leads potentially to the emergence of ghost-like instabilities. At the perturbative level, it spoils the unitarity of the quantum field theory \cite{Hawking, Antoniadis}. At the classical one, it entails the arising of Ostrogradsky instabilities \cite{Ostrogradsky}. A careful investigation of the number of the degrees of freedom by using a canonical analysis has been carried out for some higher-curvature gravity theories \cite{Crisostomi:2017ugk}, including $f(\,^{*}RR)$ ones, where it was found that these theories possess at least one ghost-like mode. However, these difficulties are circumvented by following the methodology employed in effective field theories (EFT) \cite{Georgi}, in which the ghost-like modes can be disregarded since they are suppressed by a typical high energy scale, rendering them heavy modes to propagate at the low energy limit. From now on, we shall treat the generalized Chern-Pontryagin models as EFTs. 

  It is worth stressing out that by taking the divergence of Eq.\eqref{fe}\footnote{For the pure $f(R)$ case this calculation was done in \cite{DeFelice}.}, we obtain the following constraint equation,
 \begin{equation}
     \nabla_{\mu}C^{\mu\nu}=\frac{1}{2}f_{\,^{*}RR}\nabla^{\nu}\,{}^*RR,
     \label{PontryaginEq}
 \end{equation}
where we have used the Bianchi identity, $\nabla^{\mu}G_{\mu\nu}=0$, and the conservation of the stress-energy tensor, $\nabla^{\mu}T^{(m)}_{\mu\nu}=0$, to find the aforementioned equation. Note that the divergence of the term in brackets in the r.h.s. of Eq.\eqref{ctensor} resembles the Pontryagin constraint in the Chern-Simons modified theory of gravity (see f.e. \cite{Grumiller}).
Another striking relation is obtained by tracing  Eq. \eqref{eq44}. By doing so, we find
\begin{equation}
    3\Box f_{R}+Rf_{R}-2f+2\,^{*}RR f_{\,^{*}RR}=\kappa^2 T^{(m)},
    \label{traceeq}
\end{equation}
where $T^{(m)}=g^{\mu\nu}T_{\mu\nu}^{(m)}$. The previous equation depends on the Chern-Pontryagin density, in other words, the trace equation \eqref{traceeq} holds the same of $f(R)$-theories plus an additional term proportional to $\,^{*}RR$. Such a result is interesting since this equation effectively describes a propagating scalar degree of freedom ($\Phi=f_{R}$), as we shall see in section IV in more detail; earlier this correspondence between $f(R)$ gravity and scalar-tensor gravity has been discussed in \cite{DeFelice}. For the sake of simplicity, we assume the function of the Ricci scalar and Pontryagin term to be decomposed into a sum of two functions of each of these arguments separately, $f(R,{}^*RR)=f_1(R)+f_2({}^*RR)$.

Note that the modified field equations for $f_2(^{*}RR)\propto \left(\,^{*}RR\right)^{n}$, with $n\geq 1$, reduce to the GR ones for a variety of metric classes, for example, static spherically symmetric, Friedmann-Robertson-Walker, (A)dS and G\"{o}del-type metrics, since $^{*}RR=0$, for all these metrics (actually, this condition is nothing more than the Pontryagin constraint satisfied for certain metrics once the specific relation between corresponding Newman-Penrose scalars is fulfilled, see the discussion in \cite{Grumiller}). In principle, this constraint holds for some other metrics as well. However, notice that to yield non-trivially modified field equations of motion it is necessary to have $C_{\mu\nu}\neq 0$ which leads to $\,^{*}RR\neq 0$ in particular cases. For example, just this situation occurs for the Kerr metric, as shown in \cite{Yunes}. We note also that had we considered the ``agravity-like'' theory which does not involve Einstein-Hilbert term (cf. \cite{Salvio}), so that $f_1(R)=0$, and only higher derivatives are present in the action, the metrics satisfying the Pontryagin constraint would solve the equations of motion in a trivial manner since in this case  $f_R=0$ immediately, and $f_{\,^{*}RR}=0$ if $f({}^*RR)\propto ({}^*RR)^n$, with $n> 1$. In the following section, we shall investigate perturbations of some metrics within the generalized Chern-Pontryagin gravity model.

\section{Applications}

\subsection{Perturbation of the Schwarzschild spacetime}

Since the Schwarzschild metric solves the equations of motion of the Chern-Pontryagin gravity trivially, we mean $\,^{*}RR= 0$ which leads to the vanishing of the $C$-tensor, it is natural to consider its perturbations. We will here discuss perturbations of the Schwarzschild metric within the generalized Chern-Pontryagin model, adopting  $f(R,{}^*RR)=R+\beta ({}^*RR)^2$, where $\beta$ is a coupling constant.

We focus on a particular perturbation of the Schwarzschild spacetime describing slowly rotating solutions, thus we consider the perturbed metric given by
\begin{equation}
\begin{aligned}
d s^{2}= & -\left(1-\frac{2 M}{r}\right)dt^{2}+\left(1-\frac{2 M}{r}\right)^{-1} dr^{2}+r^{2}\left[d\theta^{2}+\sin^{2} \theta(d\phi-\omega(r, \theta) dt)^{2}\right],
\label{scw}
\end{aligned}
\end{equation}
where $M$ is the mass of any spherically symmetric mass distribution and the function $\omega(r, \theta)$ is of the first order in $\epsilon \sim J / M^{2}$, here $J$ is angular momentum. In this perturbation scheme, $\epsilon$ is considered to be a small parameter. Hereafter, we take account of equations up to the first order in $\epsilon$. In this situation, the Chern-Pontryagin term for the metric \eqref{scw} up to first order in $\epsilon$ is non-zero:
\begin{eqnarray}
    {}^*RR = \frac{24M\epsilon}{r^3}\left(2\frac{\partial \omega(r,\theta)}{\partial r} \cos \theta-  \frac{\partial^2 \omega(r,\theta)}{\partial\theta \partial r} \sin \theta \right).
\end{eqnarray}

 Now, let us solve the field equations \eqref{eq44} for the line element \eqref{scw}. To begin with, the $(t,\phi)$ is the only non-vanishing component of the $C$-tensor up to first-order in $\epsilon$, so 
\begin{eqnarray}
    G_{t\phi}-T_{t\phi}^{eff}=0,
\end{eqnarray}
whose explicit form is

\begin{equation}\label{eqeisnt}
    g_{1}(r,\theta) + \left(1-\frac{2M}{r}\right) \left(r g_{2}(r,\theta) +\dfrac{a L}{r^{7}}\left[4g_{3}(r,\theta)-r g_{4}(r,\theta)\right] \right)=0,
\end{equation}
where the aforementioned functions are defined as follows
\begin{equation}
\begin{aligned}
g_{1}(r,\theta) =&\,\,\dfrac{\partial^{2}\omega}{\partial \theta^{2}}\sin \theta+3\dfrac{\partial \omega}{\partial \theta}\cos \theta,\\
g_{2}(r,\theta)=&\left( \dfrac{\partial^2 \omega}{\partial r^{2}}r+4 \dfrac{\partial \omega}{ \partial r}\right)\sin \theta,\\
g_{3}(r,\theta) =&\left(\dfrac{\partial^{3} \omega}{\partial \theta^{2}\partial r}-2\dfrac{\partial \omega}{\partial r}\right)\sin \theta+ 3\dfrac{\partial^{2} \omega}{\partial\theta \partial r}\cos \theta,\\
g_{4}(r,\theta) =&\left(\dfrac{\partial^{4}\omega}{\partial \theta^{2}\partial r^{2}}- 2\dfrac{\partial^{2}\omega}{\partial r^{2}}\right)\sin \theta+3\dfrac{\partial^{3} \omega}{\partial\theta \partial r^{2}}\cos \theta,
\end{aligned}
\end{equation}
and the constants are $a=1152$ and $L=\beta M^2$. The latter one is related to the mass of the black hole and the coupling constant accompanying the higher-derivative term in the action.
 Now, we solve the above field equation (\ref{eqeisnt}) by using the variable separation method, thus, let us assume $\omega(r,\theta)=R(r) \Theta(\theta)$. By doing so, one gets a set of two ordinary differential equations, namely, 

\begin{equation}
\label{radial}R^{''} =\left(1-\frac{2M}{r}\right)\left(\dfrac{a L}{r^{7}}(C-2) ( 4R^{'}+r) + r (4R^{'}-r)\right)+ C R;
\end{equation}
and
\begin{equation}\label{omeg2}
\Theta^{''}= - \dfrac{3}{\tan\theta}\Theta^{'}+ C \Theta,
\end{equation}
where the prime stands for derivative with respect to their correspondent arguments, and $C$ is the separation constant. Therefore, we will arrive at the following solution for the angular equation (\ref{omeg2})

\begin{equation}\label{hyperge}
\begin{aligned}
\Theta(\theta) =& A\,{}_2F_{1}\left(\frac{3+ \sqrt{9-4 C}}{4}, \frac{3-\sqrt{9-4 C}}{4}, \frac{1}{2}, \cos^2 \theta \right)+\\ &B\,{}_2F_1\left( \frac{5+ \sqrt{9-4 C}}{4}, \frac{5-\sqrt{9-4 C}}{4}, \frac{3}{2}, \cos^2 \theta \right) \cos \theta,
\end{aligned}
\end{equation}
\noindent
where $A$, $B$ are arbitrary constants, and ${}_2F_1$ is the hypergeometric function. The radial equation (\ref{radial}) is a non-linear differential equation,  whose analytical solution cannot be found in the general case. However, if we consider the particular case where the separation constant $C=0$ in the equations (\ref{radial}, \ref{omeg2}), we are able to find an analytical solution. In this situation, the angular and radial solutions reduce to
\begin{equation} \label{ome1}
R(r)= C_1+ C_2 \int \frac{r^4}{a L+r^8} \mathrm{~d} r,
\end{equation}
\begin{equation}\label{ome2}
\Theta(\theta)= C_3 + C_4 \, h(\theta),  
\end{equation}
where $C_{i's}$ are arbitrary constants and 
\begin{equation}
h(\theta)=  \dfrac{\cos\theta} {\sin^{2}\theta} -\ln \left(\dfrac{1-\cos\theta}{\sin \theta}\right).  
\end{equation}

To shed more light on this solution, let us make the following choice of the arbitrary functions, namely: $C_{1}=0$, $C_{2}=J$, $C_{3}=1$ and $C_{4}=L$ in Eqs. (\ref{ome1}, \ref{ome2}). Since the parameter $L$  characterizes the deviation of our model from the general relativity, it should be suppressed by a typical high energy scale; thus it is small. Hence,  we can expand (\ref{ome1}) up to the first order in $L$ to find the effects of the additive, non-Hilbert term in our action. This procedure will lead to the following approximate solution:
\begin{equation}
\omega(r,\theta)= \dfrac{J}{r^3}+\left(\dfrac{h(\theta)}{\sin^2{\theta}}\dfrac{J}{r^3}-\dfrac{a\,J}{11r^{11}}\right)\,L+ \mathcal{O}(L^{2}).  
\end{equation}

Here, the term $\dfrac{J}{r^3}$ is the frame-dragging effect contribution coming from GR, and the expression in the parentheses is the first-order result stemming from $\beta (^{*}RR)^2$ in our action. Note that the term proportional to $r^{-11}$ is negligible for large $r$. Therefore, in this asymptotic limit, one recovers GR, as expected, plus the correction of the first order in the small parameter $L=\beta M^2$. 

\subsection{Perturbative scheme for ST-homogeneous G\"{o}del-type metrics}

The space-time homogeneous (ST-homogeneous) G\"{o}del-type class of metrics \cite{RT}, introduced in a seminal paper by Rebou\c{c}as and Tiomno as a generalization of the G\"{o}del metric \cite{Godel}, represents itself as a set of space-times allowing for existence of the closed time-like curves (CTCs) for certain values of constant parameters of these metrics. By definition, such metrics in cylindrical coordinates are given by the following line element:
\begin{equation}
ds^2=-[dt+H(r)d\theta]^2+dr^2+D^2(r)d\theta^2+dz^2,
\label{Godelm}
\end{equation}
where $H(r)$ and $D(r)$ are the metric functions which must satisfy the following ST-homogeneity conditions:
\begin{equation}
\frac{H^{\prime}(r)}{D(r)}=2\omega;\quad\, \frac{D^{\prime\prime}}{D}=m^2,
\label{STh}
\end{equation}
where the prime means derivative with respect to $r$. Note also that the parameters $(m^2,\omega)$ are constants in which $-\infty<m^2<\infty$ and $\omega\neq 0$ is the vorticity. From the ST-homogeneity requirement \eqref{STh}, one can split the ST-homogeneus G\"{o}del-type metrics into three different classes, to wit:

{\it i) hyperbolic class:} $m^2>0$ and $\omega\neq 0$
\begin{equation}
    \begin{split}
        H(r)&=\frac{2\omega}{m^2}[\cosh{(mr)}-1],\\
        D(r)&=\frac{1}{m}\sinh{(mr)},
    \end{split}
\end{equation}

{\it ii) trigonometric class:} $m^2=-\mu^2<0$ and $\omega\neq 0$
\begin{equation}
    \begin{split}
        H(r)&=\frac{2\omega}{\mu^2}[1-\cos{(\mu r)}],\\
        D(r)&=\frac{1}{\mu}\sin{(\mu r)},
    \end{split}
\end{equation}

{\it iii) linear class:} $m^2=0$ and $\omega\neq 0$
\begin{equation}
    \begin{split}
        H(r)&=\omega r^2,\\
        D(r)&=r.
    \end{split}
\end{equation}

The original G\"{o}del solution \cite{Godel} represents the particular case $m^2=2\omega^2$, then it is a particular example of the hyperbolic class of ST-homogeneous G\"{o}del-type metrics. For brevity, from now on, we shall refer to ST-homogeneous G\"{o}del-type metrics simply as to G\"{o}del-type metrics. It was shown in \cite{RT} that all three classes of these metrics present at least one region ($r_1 <r< r_2$), where there exist CTCs defined by the circle $C=\{(t,r,\theta,z); t=t_0, r=r_0, \theta \in [0, 2\pi], z=z_0\}$. In the special case of the hyperbolic class corresponding to $m^2\geq 4\omega^2$, the presence of CTCs is completely avoidable. It can be seen by using the critical radius \footnote{The critical radius is defined as the maximum radius allowing for the presence of CTCs.}  ($r_c$) equation given by  
\begin{equation}
\sinh^2{\left(\frac{m r_c}{2}\right)}=\left(\frac{m^2}{4\omega^2}-1\right)^{-1}.
\end{equation}
Notice that for $m^2\geq 4\omega^2$, the above equation does not provide any real $r_c$. Earlier, the consistency of the three classes of G\"{o}del-type metrics within various modified gravity models have been discussed in a set of papers (see e.g. \cite{ourReb} for the Horava-Lifshitz case, and \cite{BD} for the Brans-Dicke (BD) case). In particular, it was shown in \cite{Godelour1,Godelour2,Godelour3} that ${}^*RR=0$ for G\"{o}del-type metrics, as a consequence the contribution of $C$-tensor for the field equations \eqref{fe} vanishes. In this subsection, we shall examine the possibility of breaking the ST-homogeneity for G\"{o}del-type metrics within the model $f(R,{}^*RR)=R+\beta (^{*}RR)^2$. To do that, we shall introduce a perturbative approach, which will be discussed in detail below. Then we can straightforwardly substitute the perturbed metric into the field equations to find the full solutions up to first order in the perturbation parameter.

We follow the same perturbation scheme employed in \cite{Alt}, where perturbations of ST-homogeneous G\"{o}del-type metrics have been considered within the non-dynamical Chern-Simons modified gravity. This perturbative approach has the feature that the perturbed functions are stationary and, as a consequence, they rely only on the coordinates $r, \theta$ and $z$. Note that the explicit dependence on $\theta$ and $z$ signs the possibility of breaking both translation invariance along $z$-direction and axial symmetry. As a result, it leads to perturbed G\"{o}del-type metrics that do not possess ST-homogeneity. We fix the ST-homogeneous G\"{o}del-type background metrics to be denoted by $g_{\mu \nu}^{(0)}$, whose line element is given by \eqref{Godelm}, while the first-order metric corrections are represented by  $\xi g_{\mu \nu}^{(1)}$, where $\xi$ is the perturbation parameter. Putting all information together, one can write the whole metric as below 
\begin{equation}
\label{xx}
g_{\mu \nu}=g_{\mu \nu}^{(0)}+\xi g_{\mu \nu}^{(1)}+\mathcal{O}\left(\xi^{2}\right).
\end{equation}
Their explicit line element, up to first order in $\xi$, takes the form 
\begin{equation}  
\begin{aligned}\label{ds}
d s^{2}= & -\left\{\left[1+\xi h_{0}(r, \theta, z)\right] d t+\left[1+\xi h_{1}(r, \theta, z)\right] H(r) d \theta\right\}^{2}+\left[1+\xi h_{2}(r, \theta, z)\right] D(r)^{2} d \theta^{2} \\
& +\left[1+\xi h_{3}(r, \theta, z)\right] d r^{2}+
\left[1+\xi h_{4}(r, \theta, z)\right] d z^{2},
\end{aligned}
\end{equation}
where the  $h_{i's}(r, \theta, z)$ stand for the perturbed metric functions, $g_{\mu \nu}^{(1)}$. As can be seen from the perturbed metric, we are restricting our analysis to metric perturbations for which $g_{\mu \nu}^{(1)}=0$ for all off-diagonal components in this coordinate basis except for $g_{t \theta}^{(1)}$. This mirrors the structure of the background metric $g_{\mu \nu}^{(0)}$, whose only nonzero components are $g_{t \theta}^{(0)}$ and the other $g_{\mu \nu}^{(0)}$ with $\mu=\nu$. The Chern-Pontryagin term for the perturbed metric \eqref{ds} up to first order in $\xi$ is given by
        \begin{eqnarray}
 ^{*}\,RR&=& 4\xi (m^{2}-4\omega^{2})\left[2\omega \frac{\partial}{\partial z}h_{0}(r,\theta,z) +2\omega
\frac{\partial}{\partial z}h_{1}(r,\theta,z)+\omega
\frac{\partial}{\partial z}h_{2}(r,\theta,z)\right. \nonumber\\
&& -\,\omega\frac{\partial}{\partial z}h_{3}(r,\theta,z)+\frac{H(r)}{D(r)}\frac{\partial^{2}}{\partial z\,
\partial r}h_{1}(r,\theta,z)-\left.\frac{H(r)}{D(r)}\frac{\partial^{2}}{\partial z\,\partial r}h_{0}(r,\theta,z)\right].
\label{eq-Pont}
\end{eqnarray}

Upon a simple manipulation, the previous equation can be set in more conveniently, namely,
\begin{equation}
^{*}\,RR= 4\xi (m^2 -4\omega^{2})\left(\frac{H(r)}{D(r)}\frac{\partial^{2}}{\partial z\,\partial r}P(r,\theta,z)+\omega
\frac{\partial}{\partial z}Q(r,\theta,z)\right),
\label{HD}
\end{equation}
where 
\begin{eqnarray}
P(r,\theta,z)&\equiv& h_{1}(r,\theta,z)-h_{0}(r,\theta,z);\\
Q(r,\theta,z)&\equiv& 2\left[h_{0}(r,\theta,z)+h_{1}(r,\theta,z)\right]+h_{2}(r,\theta,z)-h_{3}(r,\theta,z).
\end{eqnarray}

Now, we solve the first-order perturbed modified Einstein equation \eqref{fe} for the particular model $f({}^*RR)=R+\beta (^{*}RR)^2$. To start with, let us consider the $(t,z)$-component
\begin{equation}
 \dfrac{\partial^2 }{\partial z \partial\theta}h_1(r,\theta,z)- \dfrac {\partial^2}{\partial z \partial \theta}h_0(r,\theta,z)=0, 
 \label{11}
\end{equation}
which entails $h_{0}(r,\theta,z)=h_{1}(r,\theta,z)$. Following the same procedure, the $(\theta, z)$-component 
\begin{equation}
\dfrac{\partial^2 }{\partial z \partial \theta}h_3(r,\theta,z)+2 \dfrac{\partial^2 }{\partial z \partial \theta}h_1(r,\theta,z)=0,    \label{22}
\end{equation}
leads to the constraint equation $h_3(r,\theta,z)= -2 h_1(r, \theta,z)$. Imposing this latter constraint, we find the following $(t,r)$-component of the modified Einstein equations, that is,
\begin{equation}
\begin{aligned}
&
\left(8\,\frac{\partial }
{\partial\theta}h_1 (r,\theta,z) - \frac{\partial }{\partial \theta}h_2 (r,\theta,z)\right)+\\&
16(4\,\omega^2-m^2)^2 \left(6\,\frac{\partial^3 }{\partial z^2 \partial \theta}h_1 ( r,\theta,z)-\frac {\partial^3 }{\partial z^2\partial \theta}h_2 ( r,\theta,z)\right)=0.
 \end{aligned}
 \label{33}
\end{equation}
Requiring $m^2\neq 4\omega^2$, we find that
the only solution is the trivial one, $h_1 (r,\theta,z)=h_2 (r,\theta,z)=h_3 (r,\theta,z)=0$.

It should be realized that Eqs. \eqref{11}, \eqref{22} and \eqref{33} are automatically satisfied by considering particular metric perturbations that depend only on either the $(r,\theta)$ or $(r,z) $ coordinates. The former case \footnote{Physically speaking, this case means that the metric perturbations preserve the axial symmetry along the $z$-direction.}, $h_i (r,\theta,z)=h_i (r,\theta)$, is trivial since it leads to the vanishing of the Chern-Pontryagin term \eqref{eq-Pont}, which in turn results in the vanishing of the $C$-tensor, thereby enforcing the field equations to reduce to those of GR. As for the latter case \footnote{In this case, the metric perturbations break the axial symmetry along the $z$-direction.}, $h_i (r,\theta,z)=h_i (r,z)$, it requires a more careful inspection of the remaining field equations. In particular, the $(r,z)-$component of the modified Einstein equations is given by
\begin{eqnarray}
   && \nonumber\frac{\partial }
{\partial z}h_2 (r,z)\frac{d D(r)}{dr} -\frac{\partial }
{\partial z}h_3 (r,z)\frac{d D(r)}{dr}+2\omega H(r)\frac{\partial }
{\partial z}h_0 (r,z)-2\omega H(r)\frac{\partial }
{\partial z}h_1 (r,z)+\\
&+&2  D(r)\frac{\partial^2 }
{\partial z\partial r}h_0 (r,z)+D(r)\frac{\partial^2 }
{\partial z\partial r}h_2 (r,z)=0.
\end{eqnarray}
A particular solution is found by imposing the following relationships among perturbed metric functions, $h_0 (r,z)=h_1 (r,z), h_2 (r,z)=h_3 (r,z)=-2h_1 (r,z)$. The $zz-$component 
\begin{equation}
    2h_{1}(r,z)(m^2-3\omega^2)=0,
\end{equation}
is an algebraic equation which the only non-trivial solution is  $h_{1}(r,z)=0$. 

Thus, one concludes that the particular generalized Pontryagin model $f(R,{}^*RR)=R+\beta (^{*}RR)^2$ does not support first-order perturbations of G\"{o}del-type metrics discussed above. In contrast, the Chern-Simons modified gravity exhibits non-trivial first-order contributions \cite{Alt}.

 Another important example to consider is a generalization of the Starobinsky model, which corresponds to taking $f(R,{}^*RR)=R+\alpha R^2+\beta (^{*}RR)^2$. In this situation, we proceed similarly to the previous case to solve the field equations.  Nevertheless, we impose a simplification, as the field equations become cumbersome to tackle due to the presence of higher-order derivative terms. In particular, we consider only perturbed metric functions that do not explicitly depend on the coordinate $\theta$, that is, $h_i(r,\theta,z)=h_i(r,z)$. Recalling that the choice  $h_i(r,\theta,z)=h_i(r,\theta)$ is not well-motivated because Chern-Pontryagin corrections vanish, as explained before. To begin with, let us start by considering the $(r,z)-$component of the field equations

\begin{align}
\nonumber& \Bigg[-2 D(r)\left(\frac{\partial^{4}}{\partial z \partial r^{3}} h_2 (r,z)\right)-4 D(r)\left(\frac{\partial^{4}}{\partial z \partial r^{3}} h_0 (r,z)\right)-4\left(\frac{\partial^{3}}{\partial z \partial r^{2}} h_2 (r,z)\right)\left(\frac{\mathrm{d}}{\mathrm{~d} r} D(r)\right) \\
\nonumber& -4\left(\frac{\partial^{2}}{\partial z \partial r} h_0 (r,z)\right)\left(\frac{\mathrm{d}^{2}}{\mathrm{~d} r^{2}} D(r)\right)-4\left(\frac{\partial^{3}}{\partial z \partial r^{2}} h_0 (r,z)\right)\left(\frac{\mathrm{d}}{\mathrm{~d} r} D(r)\right) \\
\nonumber& -2 D(r)\left(\frac{\partial^{4}}{\partial z^{3} \partial r} h_2(r,z)\right)+2\left(\frac{\partial^{2}}{\partial z \partial r} h_3 (r,z)\right)\left(\frac{\mathrm{d}^{2}}{\mathrm{~d} r^{2}} D(r)\right)+2\left(\frac{\partial^{3}}{\partial z \partial r^{2}} h_3(r,z)\right)\left(\frac{\mathrm{d}}{\mathrm{~d} r} D(r)\right)\\
\nonumber&-4 D(r)\left(\frac{\partial^{4}}{\partial z^{3} \partial r} h_0 (r,z)\right)-2 D(r)\left(\frac{\partial^{4}}{\partial z^{3} \partial r} h_3 (r,z)\right) \\
\nonumber& -4\left(\frac{\partial^{2}}{\partial z \partial r} h_2 (r,z)\right)\left(\frac{\mathrm{d}^{2}}{\mathrm{~d} r^{2}} D(r)\right)+4D(r)\left(\frac{\partial^{2}}{\partial z \partial r} h_0 (r,z)\right)\left(\frac{\mathrm{d}}{\mathrm{~d} r} \ln{D(r)}\right)^{2} \\
\nonumber& -4 \omega H(r)\left(\frac{\partial^{2}}{\partial z \partial r} h_1 (r,z)\right)\left(\frac{\mathrm{d}}{\mathrm{~d} r} \ln{D(r)}\right)+8 D(r) \omega^{2}\left(\frac{\partial^{2}}{\partial z \partial r} h_1 (r,z)\right) \\
\nonumber& -4\left(\frac{\mathrm{~d}}{\mathrm{~d} r} H(r)\right)\left(\frac{\partial^{2}}{\partial z \partial r} h_0 (r,z)\right) \omega-2 m^{2}\left(\frac{\partial}{\partial z} h_2 (r,z)\right)\left(\frac{\mathrm{d}}{\mathrm{~d} r} D(r)\right) \\
\nonumber& -4 \omega^{3} H(r)\left(\frac{\partial}{\partial z} h_1 (r,z)\right)+28 D(r) \omega^{2}\left(\frac{\partial^{2}}{\partial z \partial r} h_0 (r,z)\right)-2 \omega^{2}\left(\frac{\partial}{\partial z} h_3(r, z\right)\left(\frac{\mathrm{d}}{\mathrm{~d} r} D(r)\right)\\
\nonumber&+4 \omega^{3} H(r)\left(\frac{\partial}{\partial z} h_0 (r,z)\right)-4 H(r)\left(\frac{\partial^{3}}{\partial z \partial r^{2}} h_0 (r,z)\right) \omega\\
\nonumber& +2 m^{2}\left(\frac{\partial}{\partial z} h_3 (r,z)\right)\left(\frac{\mathrm{d}}{\mathrm{~d} r} D(r)\right)-14 D(r) m^{2}\left(\frac{\partial^{2}}{\partial z \partial r} h_2 (r,z)\right)+2 \omega^{2}\left(\frac{\partial}{\partial z} h_2(r, z\right)\left(\frac{\mathrm{d}}{\mathrm{~d} r} D(r)\right)\\
\nonumber&+4 H(r)\left(\frac{\partial^{2}}{\partial z \partial r} h_0 (r,z)\right) \omega\left(\frac{\mathrm{~d}}{\mathrm{~d} r} \ln{D(r)}\right)-4 m^{2} H(r) \omega\left(\frac{\partial}{\partial z} h_0 (r,z)\right)\\
\nonumber& +4 m^{2} H(r)\left(\frac{\partial}{\partial z} h_1 (r,z)\right) \omega-2\left(\frac{\partial^{2}}{\partial z \partial r} h_3 (r,z)\right)\left(\frac{\mathrm{d}}{\mathrm{~d} r} \ln{D(r)}\right)^{2}D(r) \\
\nonumber& +4\left(\frac{\partial^{2}}{\partial z \partial r} h_2 (r,z)\right)\left(\frac{\mathrm{d}}{\mathrm{~d} r} \ln{D(r)}\right)^{2}D(r)+10 D(r) \omega^{2}\left(\frac{\partial^{2}}{\partial z \partial r} h_2 (r,z)\right) \\
\nonumber& -8 D(r) m^{2}\left(\frac{\partial^{2}}{\partial z \partial r} h_3 (r,z)\right)+8 D(r)\left(\frac{\partial^{2}}{\partial z \partial r} h_3 (r,z)\right) \omega^{2} \\
\nonumber& +4 \omega\left(\frac{\mathrm{~d}}{\mathrm{~d} r} H(r)\right)\left(\frac{\partial^{2}}{\partial z \partial r} h_1 (r,z)\right)-28 D(r) m^{2}\left(\frac{\partial^{2}}{\partial z \partial r} h_0 (r,z)\right) \\
\nonumber& +4 \omega H(r)\left(\frac{\partial^{3}}{\partial z \partial r^{2}} h_1 (r,z)\right)\Bigg] \alpha+H(r)\left(\frac{\partial}{\partial z} h_1 (r,z)\right) \omega+\frac{1}{2}\left(\frac{\partial}{\partial z} h_3 (r,z)(r,z)\right)\left(\frac{\mathrm{d}}{\mathrm{~d} r} D(r)\right)\\
\nonumber&-\frac{1}{2}\left(\frac{\partial}{\partial z} h_2 (r,z)\right)\left(\frac{\mathrm{d}}{\mathrm{~d} r} D(r)\right)-\left(\frac{\partial^{2}}{\partial z \partial r} h_0 (r,z)\right) D(r) \\
& -H(r)\left(\frac{\partial}{\partial z} h_0 (r,z)\right) \omega-\frac{1}{2}\left(\frac{\partial^{2}}{\partial z \partial r} h_2 (r,z)\right) D(r)=0.
\label{rz}
\end{align}
Note that this is a high-order derivative partial differential equation (PDE). The particular relationship among the perturbed metric functions, $h_0 (r,z)=h_1 (r,z), h_2 (r,z)=h_3 (r,z)=-2h_1 (r,z)$ and $h_1 (r,z)=F(r)+G(z)$,  solves Eq.\eqref{rz}. Although we have obtained a simple solution for the $(r,z)-$component, this solution does not satisfy the other components of the field equations. Indeed,  the more natural conclusion is that the field equations present a highly overconstrained set of PDEs, whose only possible solution is the trivial one., i.e,  $h_0(r,z)=h_1(r,z)=h_2(r,z)=h_3(r,z)=0$.

 Therefore, one concludes that by imposing the axial symmetry breaking along the $z$-direction, $f(R,{}^*RR)=R+\alpha R^2+\beta (^{*}RR)^2$  does not support first-order perturbations of G\"{o}del-type metrics.

\section{Generating the Chern-Simons term}

We start this section by showing that generalized Chern-Pontryagin models admit both Jordan and Einstein frame representations similar to what happens with $f(R)$ theories \cite{DeFelice}.

\subsection { Jordan frame representation}

To see how the Jordan frame representation emerges in more details, it is useful to observe that the action \eqref{eq1} can be cast into the following form
\begin{equation}
   S=\frac{1}{2\kappa^2}\int d^{4}x\sqrt{-g}\left[f(\sigma,\chi)+\frac{\partial f}{\partial \sigma}\left(R-\sigma\right)+\frac{\partial f}{\partial \chi}\left(\,^{*}RR-\chi\right)\right]+S_{m}(g_{\mu\nu},\Psi), 
   \label{eq2}
\end{equation}
where $\chi$ and $\sigma$ are auxiliary scalar fields. Varying Eq.(\ref{eq2}) with respect to $\sigma$ and $\chi$, respectively, we get
\begin{eqnarray}
\label{eq3}\frac{\partial^2 f}{\partial \sigma^2}(R-\sigma)+\frac{\partial^2 f}{\partial \sigma \partial \chi}\left(\,^{*}RR-\chi\right)&=&0,\\
\label{eq4}\frac{\partial^2 f}{\partial \chi^2}(\,^{*}RR-\chi)+\frac{\partial^2 f}{\partial \sigma \partial \chi}\left(R-\sigma\right)&=&0,
\end{eqnarray}
whose solution is simply given by
\begin{eqnarray}
    R=\sigma\,\, \mbox{and } \,\, \,^{*}RR=\chi.
    \label{re1}
\end{eqnarray}
Now, inserting Eq. (\ref{re1}) into Eq.(\ref{eq2}), we turn back into Eq. (\ref{eq1}).

Requiring the following condition 
\begin{eqnarray}
   \frac{\partial^2 f}{\partial \sigma^2}\frac{\partial^2 f}{\partial \chi^2}-\left(\frac{\partial^2 f}{\partial \sigma \partial \chi}\right)^2 \neq 0.
   \label{re}
\end{eqnarray}
and defining the quantities 
\begin{eqnarray}
    \frac{\partial f}{\partial \chi}\equiv \vartheta \quad \mbox{and} \quad \frac{\partial f}{\partial \sigma}\equiv \Phi, 
\end{eqnarray}
one can rewrite Eq.(\ref{eq2}) as a two-field scalar-tensor theory (Jordan frame), i.e.,
\begin{equation}
   S=\int d^{4}x\sqrt{-g}\left[ \frac{1}{2\kappa^2}\Phi R+\frac{1}{2\kappa^2}\vartheta \,^{*}RR-V(\vartheta,\Phi)\right]+S_{m}(g_{\mu\nu},\Psi),  
   \label{eq6}
\end{equation}
where 
\begin{equation}
    V(\vartheta,\Phi)=\frac{1}{2\kappa^2}\left[\vartheta\chi(\vartheta)+\Phi\sigma(\Phi)-f(\vartheta,\Phi)\right]
\end{equation}
is the interacting potential between both scalar fields. Note that the first and last terms in the gravity sector of the action (\ref{eq6}) resemble a BD action with the difference that the potential is composed of the BD scalar $\Phi$ field (scalaron), in addition to the scalar field $\vartheta$. The second term in Eq. (\ref{eq6}) is just the topological Chern-Pontryagin term non-minimally coupled with $\vartheta$ (this term is frequently called the Chern-Simons one, see f.e. \cite{JaPi}). Therefore, starting from the action \eqref{eq1},  we find a natural manner, different from the original proposal \cite{JaPi}, to generate the Chern-Simons term, with $\vartheta$ playing the role of the CS coupling field \cite{Yunes} and being one of the two scalar degrees of freedom coming up in the scalar-tensor representation. In this sense, the action of the model written in the Jordan frame representation can be thought of as a generalization of the Chern-Simons modified gravity. Note, however, that the previous choice, $f(R,\,^{*}RR)=R+\beta (\,^{*}RR)^2$, does not admit a two-field scalar-tensor representation since the on-shell condition \eqref{re1} is not ensured because \eqref{re} does not hold.  

\subsubsection{ The field equations in the Jordan frame representation}

 By varying the action \eqref{eq6} with respect to the metric and scalar fields:  $g_{\mu\nu}$, $\Phi$ and $\vartheta$, we are able to find the following field equations in the Jordan frame representation, namely,
\begin{eqnarray}
    \label{kk}G_{\mu\nu}&=&\frac{\kappa^2}{\Phi}(T^{(m)}_{\mu\nu}-g_{\mu\nu}V)+\frac{1}{\Phi}\left(\nabla_{\mu}\partial_{\nu}\Phi-g_{\mu\nu}\Box\Phi\right)-\frac{4}{\Phi}C_{\mu\nu};\\
    \label{kkk}R&=&2\kappa^2\frac{\partial V}{\partial \Phi};\\
    \label{kkkk}\,^{*}RR&=&2\kappa^2 \frac{\partial V}{\partial \vartheta},
\end{eqnarray}
where we defined the Cotton tensor as
\begin{equation}
C_{\mu\nu}=v_{\alpha}\epsilon^{\alpha}_{\,\,\,\beta\sigma_(\mu}\nabla^{\sigma}R^{\beta}_
{\,\,\,\nu)}+ v_{\alpha\beta}\,^{*}R^{\alpha\,\,\,\,\,\,\,\,\,\beta}_{\,\,\,(\mu\nu)},
\end{equation}
with $v_{\alpha}=\nabla_{\alpha}\vartheta$ and $v_{\alpha\beta}=\nabla_{\alpha}v_{\beta}$. Tracing Eq.\eqref{kk}, one finds
\begin{equation}
    \Box \Phi=\frac{1}{3}\Phi R+\frac{1}{3}\kappa^2 (T^{(m)}-4V).
\end{equation}
Now, combining it with Eq.\eqref{kkk}, we get
\begin{equation}
    \label{dys}\Box \Phi=\frac{1}{3}\Phi \frac{\partial V}{\partial \Phi}+\frac{1}{3}\kappa^2 (T^{(m)}-4V),
\end{equation}
which means that the scalaron fulfills a dynamical field equation that depends on the trace of the stress-energy tensor and also the interacting potential between both scalar fields. On the other hand, the CS field $\vartheta$ satisfies a non-dynamical equation \eqref{kkkk}, whose solution provides an equation $\vartheta=\vartheta(\Phi)$, and then $\vartheta$ can be completely factored out (on-shell) from Eq.\eqref{eq6}. Therefore, there exists only one scalar propagating degree of freedom in the theory.

We now take the model $f(R,\,^{*}RR)=R+\beta (\,^{*}RR)^2$ (the same one considered in the last section). In this case, the correspondent potential is simply given by
\begin{eqnarray}
    V(\vartheta)=\frac{\vartheta^2}{8\beta\kappa^2},
\end{eqnarray}
while the scalaron reduces to $\Phi=1$. By plugging this potential into Eq.\eqref{kkkk}, one finds
\begin{equation}
    \vartheta=2\beta (\,^{*}RR).\label{op}
\end{equation}
which can be integrated out of Eq.\eqref{kk} and, as a result, obtaining the field equations \eqref{fe} for the aforementioned particular model. Therefore, as expected, one concludes that the field equations of the generalized Chern-Pontryagin gravity in the Jordan frame are equivalent to those in the standard representation.

\subsection{ Einstein frame representation}

To the best of our knowledge, let us consider the conformal transformation of the metric,
\begin{equation}
\tilde{g}_{\mu\nu}=\Omega^{2}g_{\mu\nu},
    \label{conf}
\end{equation}
where the tilde describes geometrical quantities in the Einstein frame and $\Omega$ is the conformal factor (we note that this transformation is consistent only if $\partial f/\partial R>0$).  The Ricci scalar in both frames is related to each other by 
\begin{equation}
R=\Omega^{2}\left[\tilde{R}+6\tilde{\Box}\ln{\Omega}-6\tilde{g}^{\mu\nu}\tilde{\nabla}_{\mu}\ln{\Omega}\tilde{\nabla}_{\nu}
\ln{\Omega}\right].
\end{equation}
The Chern-Pontryagin term under (\ref{conf}) transforms as
\begin{equation}
    \,^{*}RR=\Omega^{4}\,^{*}\tilde{R}\tilde{R}.
    \end{equation}
Using the above expressions and picking $\Omega^2 =\Phi$, the action (\ref{eq6}), in the Einstein frame, looks like
\begin{eqnarray}
    \nonumber\tilde{S}&=&\frac{1}{2\kappa^2}\int d^{4}x\sqrt{-\tilde{g}}\left[\tilde{R}-\frac{3}{2}\tilde{g}^{\mu\nu}\tilde{\nabla}_{\mu}\ln{\Phi}\tilde{\nabla}_{\nu}\ln{\Phi}+\vartheta \,^{*}\tilde{R}\tilde{R}-\frac{2\kappa^2}{\Phi^2}V(\vartheta,\Phi)\right]+\\
    &+&S_{m}(\Phi^{-1}\tilde{g}_{\mu\nu},\Psi),
\end{eqnarray}
defining $\Phi=e^{\sqrt{\frac{2}{3}}\kappa \psi}$, the former equation becomes
\begin{eqnarray}
    \nonumber\tilde{S}&=&\int d^{4}x\sqrt{-\tilde{g}}\left[\frac{1}{2\kappa^2}\tilde{R}-\frac{1}{2}\tilde{g}^{\mu\nu}\tilde{\nabla}_{\mu}\psi\tilde{\nabla}_{\nu}\psi+\frac{1}{2\kappa^2}\vartheta \,^{*}\tilde{R}\tilde{R}-
    U(\vartheta,\psi)\right]+\\
    &+&S_{m}(e^{-\sqrt{\frac{2}{3}}\kappa \psi}\tilde{g}_{\mu\nu},\Psi)
    \label{eqww}
\end{eqnarray}
where the potential is
\begin{equation}
    U(\vartheta,\psi)=\frac{1}{2\kappa^2 e^{2\sqrt{\frac{2}{3}}\kappa \psi}}\left[\vartheta\chi(\vartheta)+e^{\sqrt{\frac{2}{3}}\kappa \psi}\sigma(\psi)-f(\vartheta,\psi)\right].
\end{equation}
The action in the shape of \eqref{eqww} means the Einstein frame representation. It can be easily seen that $\psi$ is the only propagating scalar degree of freedom while the CS coupling field is a non-propagating one, similarly to the original proposal of the CS modified gravity \cite{JaPi}.
 
 Let us consider the particular case corresponding to $f(R,\,^{*}RR)=R+\alpha R^2 +\beta (\,^{*}RR)^{2}$ in order to illustrate the Einstein frame representation and its connection with the CS modified gravity. In this scenario, the potential for this particular model, in the Einstein frame, takes the form 
\begin{equation}
    U(\vartheta,\psi)=\frac{1}{8\kappa^2 e^{2\sqrt{\frac{2}{3}}\kappa \psi}}\left[\frac{\vartheta^2}{\beta}+\frac{1}{\alpha}\left(e^{\sqrt{\frac{2}{3}}\kappa \psi}-1\right)^2\right]
    \label{eqw}
\end{equation}
and the action looks like
\begin{eqnarray}
    \nonumber\tilde{S}&=&\frac{1}{2\kappa^2}\int d^{4}x\sqrt{-\tilde{g}}\left\{\tilde{R}+\vartheta \,^{*}\tilde{R}\tilde{R}-\frac{1}{4 e^{2\sqrt{\frac{2}{3}}\kappa \psi}}\left[\frac{\vartheta^2}{\beta}+\frac{1}{\alpha}\left(e^{\sqrt{\frac{2}{3}}\kappa \psi}-1\right)^2\right]\right\}-\\
    &-&\frac{1}{2}\tilde{g}^{\mu\nu}\tilde{\nabla}_{\mu}\psi\tilde{\nabla}_{\nu}\psi+S_{m}(e^{-\sqrt{\frac{2}{3}}\kappa \psi}\tilde{g}_{\mu\nu},\Psi),
\end{eqnarray}
which is the non-dynamical Chern-Simons modified gravity plus a non-trivial Lagrangian depending on two interacting scalar fields.

\vspace*{3mm}

\subsubsection{ The field equations in the Einstein frame}

From now on, for the sake of convenience, we will work with the Einstein frame representation of $f(R,\,^{*}RR)$ theories.  Upon varying the action (\ref{eqww}) with respect to $\tilde{g}_{\mu\nu}$, $\psi$ and  $\vartheta$, respectively, we find the following field equations
\begin{eqnarray}
\tilde{G}_{\mu\nu}+4\tilde{C}_{\mu\nu}&=&\kappa^{2}\left(\tilde{T}_{\mu\nu}^{(\psi,\vartheta)}+\tilde{T}_{\mu\nu}^{(m)}\right);\\
  \frac{\,^{*}\tilde{R}\tilde{R}}{2\kappa^2}&=&-\frac{\partial U}{\partial \vartheta};\\
  \tilde{\Box}\psi&=&\frac{\partial U}{\partial \psi}+\sqrt{\frac{1}{6}}\kappa \tilde{T}^{(m)},
\end{eqnarray}
where we have defined the following quantities: the Cotton tensor
\begin{equation}
\tilde{C}_{\mu\nu}=\tilde{v}_{\alpha}\tilde{\epsilon}^{\alpha}_{\,\,\,\beta\sigma_(\mu}\tilde{\nabla}^{\sigma}\tilde{R}^{\beta}_
{\,\,\,\nu)}+ \tilde{v}_{\alpha\beta}\,^{*}\tilde{R}^{\alpha\,\,\,\,\,\,\,\,\,\beta}_{\,\,\,(\mu\nu)},
\end{equation}
with $\tilde{v}_{\alpha}=\tilde{\nabla}_{\alpha}\vartheta$ and $\tilde{v}_{\alpha\beta}=\tilde{\nabla}_{\alpha}v_{\beta}$. The stress-energy tensor of the matter sources is defined as usual
\begin{equation}
    \tilde{T}_{\mu\nu}^{(m)}=-\frac{2}{\sqrt{-\tilde{g}}}\frac{\delta \left[S_{m}(e^{-\sqrt{\frac{2}{3}}\kappa \psi}\tilde{g}_{\mu\nu},\Psi)\right]}{\delta \tilde{g}^{\mu\nu}},
\end{equation}
as a result, $\tilde{T}^{(m)}\equiv\tilde{g}^{\mu\nu}\tilde{T}_{\mu\nu}^{(m)}$. The stress-energy tensor
\begin{equation}
\tilde{T}_{\mu\nu}^{(\psi,\vartheta)}=\tilde{\nabla}_{\mu}\psi\tilde{\nabla}_{\nu}\psi -\tilde{g}_{\mu\nu}\left(\frac{1}{2}\tilde{\nabla}_{\alpha}\psi\tilde{\nabla}^{\alpha}\psi + U(\vartheta,\psi)\right),
\end{equation}
arises from the exclusive contributions of the scalar fields $\vartheta$ and $\psi$. Having obtained the field equations, our intention is primarily to investigate the possible solutions to this theory. For this, we concentrate on two different types of backgrounds, namely, static spherically symmetric and G\"{o}del-type metrics. For the former metric, the solutions are trivial since they lead to $\,^{*}\tilde{R}\tilde{R}=0$ and $\tilde{C}_{\mu\nu}=0$, while, for the latter one, the solution is non-trivial since it results in $\,^{*}\tilde{R}\tilde{R}=0$ and $\tilde{C}_{\mu\nu}\neq 0$. 

\vspace*{3mm}

\subsubsection{G\"{o}del-type vacuum solutions in the presence of a trivial potential $U=0$}

 In the case of a trivial potential, the field equations (added by a cosmological constant) reduce to a similar shape to those found in \cite{Godelour2} unless by the fact that, here, $\tilde{g}$ is the dynamical metric. By solving the field equations, the authors of \cite{Godelour2} found the following solutions for the scalar fields: $\vartheta=b(z-z_0)$ and $\psi=s(z-z_0)$, where $z_0$ is an arbitrary real constant. Therefore, formally, the case corresponding to the trivial potential, $U=0$, just reduces to those found for CSMG \cite{Godelour1,Godelour2,Godelour3}, as expected.

\section{Summary}

We introduced a new modified gravity model called generalized Chern-Pontryagin models, $f(R,{}^*RR)$ ones, where the additive term is a function of the scalar curvature and the Chern-Pontryagin invariant ${}^*RR$. The interest to this model is motivated by the fact that it, being the generalization of CSMG, allows for parity breaking in certain situations, that is, when the Lagrangian involves odd degrees of ${}^*RR$, which motivates further applications of the model within studies of CPT-Lorentz breaking context.

It is easy to check that many known metrics solve the equations of motion for our model, namely, all those ones which yield the result ${}^*RR=0$; in this case, the field equations reduce to Einstein ones. In particular, it is true for all spherically symmetric metrics. This fact motivated us to study perturbations of two well-known metrics, namely, Schwarzschild and G\"{o}del ones. In a certain sense, our study can be treated as a continuation of the paper \cite{Alt} where the perturbations of G\"{o}del-type metrics have been considered within the non-dynamical CSMG. We investigated the consistency of these metrics within the particular models $f(R,{}^*RR)=R+\beta ({}^*RR)^2$ and $f(R,{}^*RR)=R+\alpha R^2 +\beta ({}^*RR)^2$. For the perturbations of the Schwarzschild black hole, we have found a first-order frame-dragging effect in the perturbation parameter $L=\beta M^2$, apart from the standard GR term. For the perturbations of the G\"{o}del-type metrics, we have investigated both general first-order perturbations and those displaying axial symmetry breaking along the $z-$direction. We conclude that, in these scenarios, all perturbed metric functions must equal zero to ensure consistency with the field equations.

Similar to $f(R)$ theories, generalized Chern-Pontryagin models admit a two-field scalar-tensor representation with an interacting potential of such two scalar fields 
and, as a consequence, Jordan and Einstein frames. In this representation, it is clear the link between these models and Chern-Simons modified gravity. Actually, by describing the theory in the Einstein frame, we have shown that the action of this class of models is to some extent an intermediate case between the non-dynamical and dynamical CSMG. Furthermore, we have succeeded in engendering the Chern-Simons term more naturally than in the original proposal of CSMG \cite{JaPi}.

The natural continuation of this study could involve investigating the study of cosmological perturbations and also the implications of the parity violation in the context of gravitational waves. We plan to address these issues in the forthcoming paper.


{\bf Acknowledgments.} 
 This work was supported by Conselho Nacional de Desenvolvimento Cient\'{\i}fico e Tecnol\'{o}gico (CNPq) and Paraiba State Research Foundation (FAPESQ-PB). PJP would like to thank the Brazilian agency CNPq for financial support (PQ--2 grant, process 
 No. 307628/2022-1).  The work by AYP has been supported by the CNPq project No. 303777/2023-0. PJP, AYP and JRN  thank the Brazilian agency FAPESQ-PB for financial support (process No. 150891/2023-7).




\end{document}